\documentclass[pdflatex,sn-mathphys-num, iicol]{sn-jnl}
\usepackage{graphicx}
\usepackage{color}
\usepackage{bm}
\usepackage{bbm}
\usepackage{amsmath}
\usepackage{amssymb}
\usepackage[normalem]{ulem}
\usepackage{stackengine}

\def\gtrless{\raise2.5pt\hbox{$>$}\llap{\lower2.5pt\hbox{$<$}}}
\def\gtrapprox{\raise2.5pt\hbox{$>$}\llap{\lower2.5pt\hbox{$\approx$}}}

\newcommand{\bsq}[1]{\begin{subequations}\label{#1}}
\newcommand{\esq}{\end{subequations}}
\newcommand{\beq}[1]{\begin{equation}\label{#1}}
\newcommand{\eeq}{\end{equation}}
\newcommand{\beqa}[1]{\begin{eqnarray}\label{#1}}
\newcommand{\eeqa}{\end{eqnarray}}

\newcommand{\mb}{\mathbf}

\renewcommand{\rho}{\varrho}
\renewcommand{\epsilon}{\varepsilon}
\graphicspath{{./EPJfigures/}{./EPJfigure/3dplots}}

\begin{document}

\title{Amoeboid swimming of active vesicles}

\author*[1]{\fnm{Reiner} \sur{Kree}}\email{rkree1@phys.uni-goettingen.de}\equalcont{These authors contributed equally to this work.}
\author[1]{\fnm{Annette} \sur{Zippelius}}
\equalcont{These authors contributed equally to this work.}

\affil[1]{\orgdiv{Inst. Theoret. Physics}, \orgname{Georg-August University G\"ottingen}, \orgaddress{\street{Friedrich-Hund Pl. 1}, \city{G\"ottingen}, \postcode{D-37077}, \country{Germany}}}
\date{\today}

\abstract{
We investigate the shape dynamics and migration of weakly deflated
active vesicles driven by processes acting either directly in the
membrane or transmitted by the cytoskeleton. For a force-free vesicle,
local membrane incompressibility suppresses rigid-body translation, so
that migration arises from time-dependent shape deformations.
Assuming small excess area enables a systematic analysis of the
coupled deformation and migration dynamics in free space, i.e. in the
absence of substrate adhesion or confinement. Depending on the
strength and frequency of the activity, the vesicle exhibits several
 dynamical regimes, including synchronized oscillations,
quasiperiodic shape changes, transitions between non-propelling and
propelling states, and intermittent motion.
 }

\maketitle

\section{Introduction}
Microorganisms have developed various strategies to migrate in complex environments~\cite{Othmer2019}.
One such mechanism relies on active shape deformations generated by forces within the cell. 
In the absence of adhesion to the substrate or confinement, these deformations are coupled to the surrounding fluid through viscous stresses and can generate locomotion. The resulting motion, called
amoeboid swimming, has only recently been recognized as a possible
migration mode for cells~\cite{Paluch2016,Bergert,Jones,Noselli},
which have been placed in solution and were observed to swim \textit{in
    vitro}.  Such cells include Dictyostelium amoebae, neutrophils~\cite{Garcia_Seyda}, lymphocytes, and other cell types~\cite{Barry,ONEILL2018,Bodenschatz,AOUN2020}, and has even been reported \textit{in vivo} for fat body cells in Drosophila~\cite{Andrieu}.

Cellular motility has also motivated the development of artificial cells and active soft materials ranging from autonomous systems to externally controlled ones. Vesicles containing active components are among the simplest realizations of such systems. Their deformability and shape changes depend sensitively on the properties of the enclosing membrane, which can be controlled in several ways~\cite{McMahon2005,Dimova2014}. 
For example, photoswitchable lipids in giant unilamellar vesicles (GUVs) allow reversible shape changes with high temporal and spatial resolution~\cite{Pernpeintner2017_Langmuir,Pritzl2025_Review}. 
Active membrane proteins provide another mechanism for generating dynamic shape changes.
The MIN system in GUVs was shown to form moving
protein-density waves on the membrane~\cite{Litschel2018}. The
accompanying periodic and reversible shape changes of the vesicles have been attributed to protein-induced spontaneous curvature~\cite{Christ2021}. 
Similar mechanisms can be experimentally realized by optogenetic control of the membrane curvature. In particular, engineered BAR-domain proteins allow light-induced generation of positive and negative membrane curvature even in living cells~\cite{Jones2020}. 

Theoretical approaches to amoeboid swimming are mainly based on active
vesicles in Stokes flow. Several groups have simulated two-dimensional
soft vesicles, filled with active
particles~\cite{Paoluzzi2016},\cite{Tian2017}. Paoluzzi et
al.~\cite{Paoluzzi2016} observe deformations for intermediate
concentrations of active particles, while the vesicles perform
time-persistent random walks, reminiscent of Eukaryotic cells. Abaurrea et
al.~\cite{AbaurreaVelasco2019} consider vesicles containing active
filaments and demonstrate self-organized propulsion and complex shape
transformations. Iyer et al.~\cite{Iyer2022} study vesicles filled
with active Brownian particles and observe various nonequilibrium
shapes. Farutin et al.~\cite{Farutin2013} model the swimmer as a
nearly spherical vesicle, which is driven by a prescribed normal
force, acting on the enclosing membrane. Time-periodic deformations of
the membrane give rise to a finite propulsion velocity. More
complex computational models~\cite{Lim2013, Campbell2017} include membrane-cortex coupling to model
swimming of blebbing cells. The control of membrane
curvature by embedded proteins has been modeled in the context of vesicle
division, using purely relaxational dynamics of the
Ginzburg-Landau type \cite{Barrio2020}.

These examples raise the question of how active processes generate shape dynamics and propulsion of vesicles.
Activity may originate in the cortex, where molecular motors generate stresses that are transmitted to the membrane and produce local traction or flow. Activity may also originate in the membrane itself, where spatial variations of protein or lipid composition modify material properties such as spontaneous curvature or bending rigidity. We describe both mechanisms by effective active tractions and determine the resulting shape dynamics from the membrane force balance. For nearly spherical vesicles, the corresponding swimming velocity follows from a systematic expansion in the excess area~\cite{Farutin2013,Kree2025}. In this way, microscopic active processes can be related to shape dynamics and autonomous propulsion.

We now proceed to the theoretical analysis and its consequences. Section~\ref{sec:model} reviews the relation between deformation and propulsion and formulates the model. Section~\ref{solution}  derives the equations governing the active shape dynamics, and Section~\ref{results} discusses their consequences for two- and three-mode driving. 
We summarize our findings in Section~\ref{discussion}. Technical details are collected in the appendices.

\section{Model}
\label{sec:model}
We consider a vesicle that is suspended in a Newtonian fluid of
viscosity $\eta$. The vesicle is enclosed by a locally inextensible fluid membrane and contains another Newtonian fluid of viscosity  $\lambda\eta$. We first review how shape deformations generate  propulsion~\cite{Lighthill,Farutin2013,Kree2025} and subsequently derive the deformations generated by membrane elasticity and active driving forces.

\subsection{Propulsion velocity}
In the absence of the vesicle, the ambient fluid is at rest in the laboratory frame. At low Reynolds number, the flow fields inside and outside the vesicle satisfy the Stokes equations
\begin{equation}
\label{eq:stokes}
\nabla\cdot\mb{\sigma}=\eta\nabla^2\mb{v}-\nabla p=0 , 
\end{equation}
supplemented by the incompressibility condition $\nabla\cdot\mb{v}=0$.
The stress tensor $\bm{\sigma}$ is given by the Cartesian components
$\sigma_{ij}=-p\delta_{ij}+\eta(\partial_iv_j+\partial_jv_i) =
-p\delta_{ij} + \sigma^{visc}_{ij}$, with pressure $p$ determined
from incompressibility.

The shape of the vesicle is described as a weakly deformed sphere; in the laboratory frame the points on its surface S have a position vector $\mb{x}$  
\begin{equation}
 \mb{x}=\mb{R}+a(1+f)\mb{e}_r,
    \label{eq:paramdeformation}
  \end{equation}
  where $R$ denotes the position of the vesicle and $f(\theta,t)$ the shape deformation with respect to a sphere
 of radius $a$ and volume $V=4\pi a^3/3$.  The deformation
  $f=f(\theta,t)=\sum_lf_lP_l(\theta)$ is taken to be axially
  symmetric and hence can be expanded in Legendre
  polynomials $P_l$. 

  The choice of the reference point $\bm R$ is not unique. Although one could use the center of mass of the vesicle, it is more convenient to choose $\bm R$ such that the $l=1$ component of the deformation vanishes. In this way, rigid translations, represented by $\bm R(t)$, are separated from genuine shape deformations. We therefore work in the frame centered at $\mb R(t)$, which will be called center-of-deformation (CoD) frame in the following.

The membrane is locally inextensible, which implies that the membrane is inextensible.
\begin{equation}\label{eq:inextensibility}
  \nabla_S\cdot \mb{v}=0
\end{equation}
where $\nabla_S$ denotes the divergence  on the deformed surface. Together with bulk incompressibility this implies

\begin{equation}
    n_i\sigma^{vis}_{ij}n_j = 0.
\end{equation} The outward normal of the surface is denoted by $\mb{n}$ and $\sigma^{vis}_{ij}$ is the viscous part of the stress tensor. Local membrane incompressibility couples normal and tangential motion on the surface. The dynamic evolution of the deformation is given by  the kinematic boundary condition
\begin{equation}
  \label{eq:BCV2}
a(\mb{e}_r\cdot\mb{n})\frac{\partial f}{\partial t}=\mb{v}\cdot\mb{n}.
\end{equation}

The propulsion velocity as a function of the deformation is determined by
the two boundary conditions
(\ref{eq:inextensibility},\ref{eq:BCV2}). It can be computed in a
perturbation expansion, assuming that the excess area $\Delta$ of the
membrane is small. More explicitly, the area of the membrane,
$A=4\pi a^2(1+\Delta)$, is constant, restricting possible deformations
\begin{align}\label{eq:excess_area}
  A&=4\pi a^2(1+\Delta)\nonumber\\
  &=2\pi a^2\int_{-1}^1 d(\cos{\theta}) (1+f)\sqrt{(1+f)^2+f'^2}.
\end{align}
To organize a perturbation theory, we introduce a small parameter $\epsilon\propto\sqrt{\Delta}$, which implies $f\propto\epsilon$. Equation(\ref{eq:excess_area}) then reads in lowest order
\begin{equation}\label{eq:area_constraint}
  \Delta=\sum_l\frac{(l-1)(l+2)}{2(2l+1)}f_l^2=:\sum_lw_lf_l^2.
\end{equation}

For nearly spherical vesicles, the swimming velocity can be expressed solely in terms of the deformation amplitudes and their time derivatives~\cite{Lighthill,Farutin2013,Kree2025}. The derivation is summarized in Appendix~\ref{appendixA}, and the resulting propulsion law reads

\begin{equation}
\label{eq:Uparticle}
  U = \sum_{l\geq 2}\big\{ C_l(f_l\dot{f}_{l+1} 
      - f_{l+1}\dot{f}_{l})+B_l\frac{d}{dt}f_lf_{l+1}\big\}
    \end{equation}
    with $C_l$ and $B_l$ given in Appendix~\ref{appendixA}. The first term is antisymmetric in neighboring modes and determines the cycle-averaged propulsion, whereas the second term is a total time derivative and does not contribute to the  velocity averaged over a periodic cycle.
The time dependence of the propulsion velocity will be discussed,
    once the time dependent deformations have been computed as a
    function of the activity. This will be the subject of the next
    subsections.
    
\subsection{Coupling activity to deformation and flow}
Shape deformations are generated by active processes in the system. In
our minimal model of a vesicle, we consider two mechanisms: 1)
Inhomogeneous and time dependent membrane properties are created
e.g. by proteins, which are associated to the membrane. 2) Actomyosin
activity in the cortex generates stresses that can be transmitted to
the membrane~\cite{Schwille2012}, leading to local contraction, tension gradients, and
flow~\cite{CallenJones}.  We model these mechanisms by coarse grained
active tractions $\mb{F}^{act}=\mb{F}^{cortex}+\mb{F}^{mem}$.

The membrane forces comprise passive and active parts and are derived from the Helfrich free energy
\begin{equation}
 {\cal{F}}=\int_S dA \big(\frac{\kappa}{2} (H-C)^2+\gamma \big)+p_0 V. 
\end{equation}
The first term accounts for the bending of the membrane with the mean
curvature $H=\nabla_S\cdot \mb{n}$ and the spontaneous curvature
$C(\mb{x})$. Both spontaneous curvature and the bending
modulus $\kappa(x)$ are in general inhomogeneous and may actively drive
shape deformations.  We do not include the Gaussian curvature. It is
important only if either topological changes are considered -- which
we do not -- or if the Gaussian bending modulus is
inhomogeneous. Limited information about the Gaussian bending modulus
is available from experiment~\cite{Holderer2013} or
simulation~\cite{HuBriguglioDeserno2012,Leermakers2013} and at best for the spatially
averaged modulus. 


 The surface energy
$\gamma(\mb{x})$ plays the role of a Lagrange parameter, ensuring
the local inextensibility of the membrane. We decompose it into a uniform
and inhomogeneous part:
$\gamma(\mb{x})=\gamma_0+\delta \gamma(\mb{x})$. The constant part ,
$\gamma_0$, enforces the global area constraint of
Eq.(\ref{eq:area_constraint}). Finally, the constant pressure $p_0$
enforces a constant volume of the vesicle.


The activity of actomyosin in the cortex is modeled as an active
stress $\mb{\sigma}^{cortex}$ that acts on the membrane. The
corresponding traction
$\mb{F}^{cortex}=\mb{n}\cdot\mb{\sigma}^{cortex}$ has both normal and
tangential components and, in general, there can be active shear as well
as active pressure. For an autonomous system, we require that the
total active force vanishes
\begin{equation}\label{eq:force_free}
  \int_S dA\;\mb{F}^{cortex}=0.
\end{equation}

Membrane deformations cause flows both in the interior and
exterior fluid, which in turn generate viscous stresses $\sigma$, acting on 
the membrane. 
The balance of force at the membrane relates the viscous tractions to the active tractions, derived from membrane forces and/or the cortex. 

\begin{equation}\label{eq:force_balance1}
  \mb{\sigma^+}\cdot\mb{n}-\mb{\sigma^-}\cdot\mb{n}=\mb{F}^{cortex}+\mb{F}^{mem}.
\end{equation}
Here, the superscript $+(-)$ denotes the exterior (interior) stress. 
The flow is continuous across the membrane , so the velocity in the interior of the vesicle $\mb{v}^-$ can be calculated from the exterior velocity $\mb{v}^+=\mb{v}^-$ on the surface $S$. The shape deformations and the associated flows are determined by this balance.

\section{Active shape dynamics}
\label{solution}
Assuming weak deviations from a sphere, we retain only linear terms in the deformation amplitudes.  To set up a perturbation theory in
$\epsilon$ we expand the mean curvature
\begin{equation}\label{mean_curvature_expansion}
H=\frac{2}{a}-\frac{\epsilon}{a}(\Delta+2)f+{\cal{O}}(\epsilon)^2
\end{equation}
and we take $C_0=2$ for simplicity. The more general case $C_0\neq 2$ is discussed in Ref. \cite{Kree2026}. 

The tractions on the right hand side of Eq.(\ref{eq:force_balance1}) then consist of passive bending forces and active tractions $\mb{F}^{act}$, due to both, cortex activity and  nonuniform elasticity  
\begin{align}
  \label{eq:normal_force}
  F_r&=\frac{\kappa}{a^3}(\Delta+2)^2f
       -\frac{\gamma_0}{a}(\Delta+2)f+\frac{2}{a}\gamma(\theta)+F_r^{act}\\
 \label{eq:tangential_force} 
  F_{\theta}&=-\frac{\gamma^{\prime}(\theta)}{a}+F_{\theta}^{act}.
\end{align}
The derivation of the active bending forces due to inhomogeneous elasticity is outlined in Appendix~\ref{appendixC}.

We expand the surface tension and the active tractions in Legendre
polynomials. Combining these tractions with the hydrodynamic tractions calculated in Appendix~\ref{appendixB} yields two coupled force-balance equations for each mode $l\ge 2$:
\begin{align}
  \label{eq:force1_main}
  -N_l\dot{f}_l&=g_l\Big(\frac{\kappa}{a^3}g_l+\frac{\gamma_0}{a}\Big)f_l+\frac{2}{a}\gamma_l +F_{r,l}^{act}\\
    \label{eq:force2_main}
  -T_l\dot{f}_l &=-\frac{\gamma_l}{a}+F_{\theta,l}^{act}
\end{align}
with $g_l=l(l+1)-2$ and $N_l,T_l$ given in Appendix~\ref{appendixB}.

The Lagrange parameter $\gamma_l$, which enforces the local inextensibility of the membrane, is determined from Eq.(\ref{eq:force2_main}) and substituted into Eq.(\ref{eq:force1_main}) to obtain the evolution equation for the deformation $f$
\begin{equation}\label{evolution_f}
  -(N_l+2T_l)\dot{f}_{l}=g_l(\frac{\gamma_0}{a}+\frac{\kappa}{a^3}g_l)f_l+F_{r,l}^{act}
  +2F_{\theta,l}^{act}.
\end{equation}
As the last step, we enforce the global constraint of a constant area (\ref{eq:area_constraint}) 
\begin{equation}
  \dot{\Delta}=0=\sum_l\frac{g_l}{2l+1}f_l\dot{f}_l.
\end{equation}
by choosing $\gamma_0$ appropriately. Thereby a closed nonlinear
evolution equation for the time-dependent deformation has been
derived. Its solution determines the flow fields and the
propulsion velocity, as will be discussed in the next section.

The polar mode ($l=1$) requires some care. Since
$g_1=0$ the passive bending tractions in Eq. (\ref{evolution_f}) vanish. Furthermore, the total active force $F_{r,l}^{act}+2F_{\theta,l}^{act}$ has to vanish. Hence, the r.h.s. of Eq. (\ref{evolution_f}) is zero.
As shown explicitly in Appendix~\ref{appendixA},  a spherical vesicle driven by $l=1$ tractions behaves exactly like a rigid sphere. A non-vanishing force $F_z\bm{e}_z$ leads to a propulsion velocity related to the force by Stokes law $F_z=6\pi\eta^+ a\, U$, implying $U=0$ for an autonomous system. 
This result holds beyond perturbation theory because it only depends on inextensibility, in other words on the absence of sinks and sources.

Hence, within the present quasi-spherical description, autonomous propulsion
arises only through nonlinear coupling between deformation modes with
$l\ge2$. For a perfectly spherical vesicle, a force-free polar
activity is completely compensated by the tension field and does not
generate motion. For a nearly spherical vesicle,
a polar pattern of active stresses or retrograde cortical flow alone
does not produce free swimming.
Shape deformations are essential for converting such activity into
locomotion.

\section{From shape dynamics to propulsion}
\label{results}
Periodic activity is a natural ingredient of many active processes, including actomyosin waves in the cortex and oscillations of protein density in the membrane. To illustrate the resulting dynamics, we focus on harmonic driving and consider two simple scenarios involving harmonic driving with either two or three adjacent modes.
Since Eq.~(\ref{evolution_f}) is nonlinear due to the global area constraint, its solution generally requires numerical methods. For this purpose, it is convenient to work with dimensionless quantities. We choose $a$ as the unit of length, $\tau_b=\eta^+a^3/\kappa$ as the unit of time, $\kappa_0/a^3$ as the unit of stress, $\kappa/a^2$ as the unit of tension, and $a/\tau_b$ as the unit of velocity. Henceforth, all variables are dimensionless.

The central questions addressed in the following are:
How does the swimming speed depend on the strength and frequency
of the activity? Under which conditions does periodic forcing
generate net propulsion? And how are the deformation cycles
synchronized with the external drive?




 \subsection{Two-mode driving}

 The simplest but nevertheless instructive case is harmonic driving with a single
 frequency $\omega=2\pi/T$, retaining only one adjacent pair of active
 driving forces $(l,l+1)$ in Eq.(\ref{evolution_f}). 
  We choose the pair $(2,3)$ for the active tractions $F_l^{act}=F_{r,l}^{act}+2F_{\theta,l}^{act}$ in Eq.(\ref{evolution_f})
 \begin{equation}\label{two_modes}
   F_2^{act}=s\cos(\omega t),\qquad F_3^{act}=\alpha s\cos(\omega t+\delta),
 \end{equation}
 where $s$ denotes the overall strength of the activity, $\delta$ the
 relative phase, and $\alpha$ the relative strength of the amplitudes.
 The generated deformations $(f_2,f_3)$ have to meet
 the excess area constraint, which is explicitly read as
 \begin{equation}
   \Delta=w_2f_2^2+w_3f_3^2. 
\end{equation}
The shape dynamics is thus confined to an ellipse. It is convenient to
rescale the deformations $q_l=f_l\sqrt{w_l/\Delta}$, so that the
motion is on the unit circle $1=q_2^2+q_3^2$. The shape dynamics is
one-dimensional, completely described by the only remaining degree of
freedom, the angle $\Psi(t)$, locating the instantaneous state of the
system on the cycle: $q_2=\cos\Psi$ and $q_3=\sin\Psi$. We thus need
to analyze motion on a circle driven by periodic forcing.

Two qualitatively different types of motion on the circle are
possible: Either the system oscillates back and forth over a finite
region of the circle, or it winds around the circle, possibly many
times. This dynamics is reminiscent of a planar pendulum, which either oscillates back and forth or runs around. If the trajectory completes a full circle in time $\tau$, then the time averaged propulsion velocity is 
\begin{equation}
    \overline{U}(\tau)=\frac{1}{\tau}\int_0^{\tau}dt^{\prime} U(t^{\prime})=\frac{2}{\tau}{\tilde{C}}_2A_{2,3}(\tau). 
    \label{timeaveragedU}
  \end{equation}
  here ${\tilde{C}}_2=C_2\Delta/\sqrt{w_2 w_3}$, and
$A_{2,3}(\tau)=\pi$ denotes the area enclosed by the trajectory in the $(q_2,q_{3})$ plane in $\tau$.  
We have used the identity
  \begin{equation}
\int_0^{\tau}dt\; \dot{q}_{3}q_2 = \oint q_2\, dq_{3}=A_{2,3}(\tau)=\pi.
 \end{equation}
 The second term in Eq.(\ref{eq:Uparticle}) does not contribute, because the end point of the trajectory coincides with the starting point.
 
 For the librating
 state, we identify $\tau$ with the time needed to return to the
 initial state. The time averaged propulsion velocity vanishes
 because the trajectory does not include a finite area.

More generally, propulsion requires
 {\it{non-reciprocal}} shape changes. If the deformation retraces
 itself in time (time-reversible motion), the enclosed area vanishes
, and no net propulsion occurs. Conversely, any cyclic deformation that
 encloses a finite area in shape space leads to directed motion.

 To analyze two-mode dynamics quantitatively, we reformulate the two
 dynamic equations~(\ref{evolution_f}) for the deformations $f_2,f_3$
 as an ordinary first order differential equation for the phase
 $\Psi(t)$, as shown in the Appendix~\ref{appendixD}. This equation has to
 be solved numerically and we discuss here only the results, leaving a
 detailed analysis to  a different paper~\cite{Kree2026}.
 
\begin{figure}[htbp]
    \centering
    \raisebox{0.7 cm}{\stackinset{l}{1pt}{t}{-20pt}{\textbf{(a)}}{
    \includegraphics[width=0.18\textwidth]{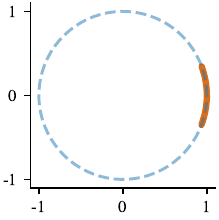}}}
    \hfill
    \stackinset{l}{30pt}{t}{-10pt}{\textbf{(b)}}{
    \includegraphics[width=0.25\textwidth]{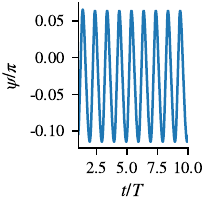}}
    \caption{\label{libration}
Non-propelling state for weak forcing in the two-mode system.
(a): the phase variable $\Psi(t)$ covers only a finite part of the
circle, corresponding to bounded shape oscillations. (b): the
associated lifted phase oscillates periodically without net growth,
indicating that the deformation cycle remains synchronized with the
driving and does not advance around the circle. Parameters:
$\hat s=5\sqrt{40}$ and $\omega=1.48$, $\alpha=5/\sqrt{14}$, $\delta= 0.6\pi$ .
}
\end{figure}

The most relevant parameter is $\tilde{s}=s/\sqrt{\Delta}$ , where $s$
is the strength of the forcing and $\Delta$ the excess area. The
passive system, $s=0$, has two stable fixed points $\Psi=0,\pi$ and two
unstable fixed points $\Psi=\pi/2, 3\pi/2$. The dynamics is purely
relaxational, and all initial conditions are drawn into one of the two
fixed points.  For weak driving, i.e., small $s$, $\Psi(t)$ oscillates
around one of the stable fixed points, but never completes a full circle,
see the left part of Fig.\ref{libration}. 
The frequency of the librating mode is locked to the
external driving frequency $\omega$.  The vesicle moves periodically,
as shown in the right part of Fig.\ref{libration}. The trajectory does
not enclose a finite area. Hence, there is no net motion and the
vesicle's speed, averaged over one period of the oscillation,
vanishes.

\begin{figure}[htbp]
    \centering
    \raisebox{0.4 cm}{\stackinset{l}{1pt}{t}{-17pt}{\textbf{(a)}}{
    \includegraphics[width=0.18\textwidth]{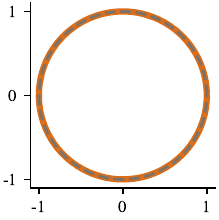}}}
    \hfill
    \stackinset{l}{30pt}{t}{-10pt}{\textbf{(b)}}{
    \includegraphics[width=0.25\textwidth]{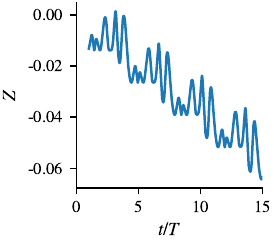}}
    \caption{\label{run} Propelling state for strong forcing in the two-mode system.
(a): the phase variable $\Psi(t)$ winds around the entire circle. (b): the resulting
displacement exhibits a monotonic drift, demonstrating directed
swimming. The superimposed oscillations reflect the periodic shape
changes. Parameters:
$\hat s=11\sqrt{40}$ and $\omega=1.48$, $\alpha=5/\sqrt{14}$, $\delta= 0.6\pi$ .}
\end{figure}

On the other hand, for large driving, $\Psi(t)$ winds around the
circle, enclosing the area of the circle, possibly many times.
The average velocity
of the vesicle is finite ~Eq.(\ref{timeaveragedU}), as seen in the
distance traveled by the vesicle (see Fig.\ref{run}).

\begin{figure}[htbp]
    \centering
    \stackinset{l}{1pt}{t}{-8pt}{\textbf{(a)}}{
    \includegraphics[width=0.2\textwidth]{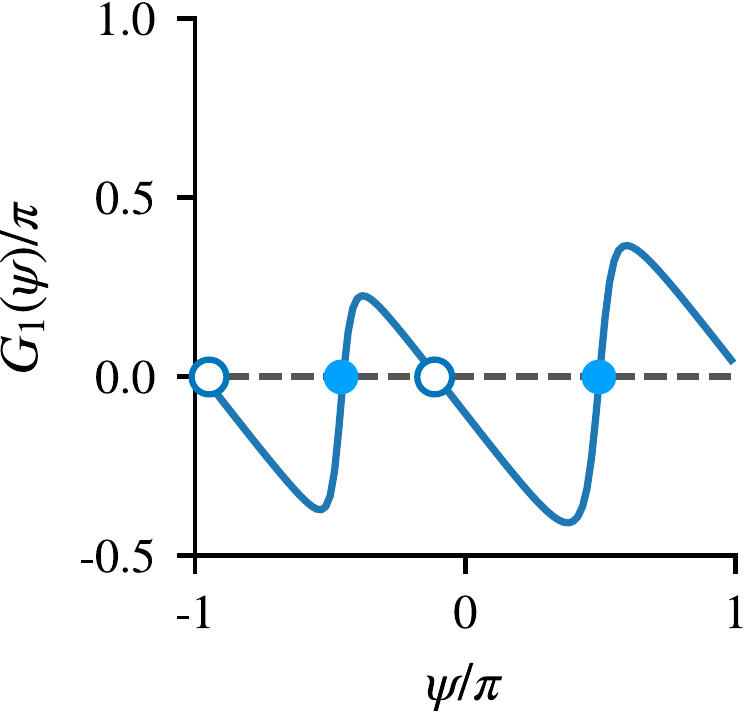}}
    \hfill
    \stackinset{l}{10pt}{t}{-10pt}{\textbf{(b)}}{
    \includegraphics[width=0.2\textwidth]
   {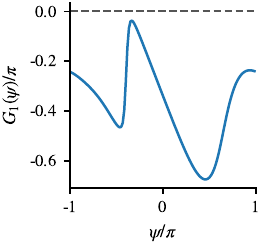}}    
    \caption{\label{stroboscopic}
Difference function $G(\psi)=P(\psi)-\psi$ for weak and strong
forcing. (a): intersections with the horizontal axis correspond to
periodic deformation cycles synchronized with the external drive.
Filled circles denote stable cycles, while open circles denote
unstable ones. (b): for strong forcing no intersections exist and
the deformation cycle progresses continuously around the circle,
resulting in directed swimming. Parameters: $\alpha=5/\sqrt{14}$, $\delta= 0.6\pi$, and
$\hat s=5\sqrt{40}$ for (a), $\hat s=11\sqrt{40}$ for (b).}
\end{figure}

To understand the transition between the two regimes, we look
at the stroboscopic map:
 \begin{equation}
     \Psi_{n+1}={\cal{P}}(\Psi_n) \qquad \Psi_n=\Psi(nT)
 \end{equation}
 A fixed point of the stroboscopic map $\Psi^*={\cal{P}}(\Psi^*)$
 implies shape dynamics, which is periodic with the same frequency as
 the driving. Such a state is called a phase locked state. In
 Fig.\ref{stroboscopic} we show the stroboscopic map for weak
 driving on the left and stronger driving on the right. Rather than plotting the map itself, we consider the difference
$
G(\psi)=P(\psi)-\psi.
$
The intersections of the graph with the horizontal axis correspond to
periodic shape cycles synchronized with the forcing. 
 For weak
 driving, we find two stable and two unstable fixed points. This implies two distinct, non-winding shape cycles, both
 synchronized with the driving, as discussed above. The system is thus multistable, and it
 depends on the initial condition, which of the two fixed points is
 reached for long times.

\begin{figure}
  \includegraphics[width=0.45\textwidth]{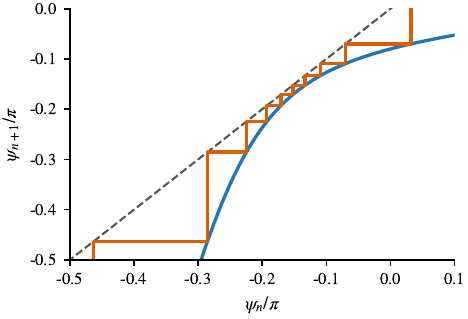}
  \caption{\label{iteration}
Cobweb construction of the stroboscopic map; it illustrates the origin of a diverging timescale at the transition to the propelling phase.  Parameters $\hat s=1.0245,\, \omega=1.48$, $\alpha=5/\sqrt{14}$, $\delta= 0.6\pi$ .
}
\end{figure}

Above a critical forcing $s^*$, the shape oscillations are no longer confined to a neighborhood of the stationary shape. The deformation cycle begins to wind around the full circle, producing net propulsion.
 Above threshold, the phase exhibits unbounded drift,
 resulting in propulsion. In Fig.\ref{iteration} we show a detail of the stroboscopic map for
a parameter value slightly above threshold and still
close to the bifurcation. 
It shows that long intervals of nearly periodic,
non-propelling oscillations are interrupted by occasional phase-slip
events that advance the deformation cycle and generate propulsion.
The dynamics takes the system close to a preferred shape cycle, where  fixpoints will be created by a further decrease of the parameter. Eventually it
escapes from this quasi-stationary state through a phase-slip event, which advances the deformation cycle and produces a burst of
propulsion. 

From Fig. \ref{iteration} it is obvious that the residence time near
the quasi-stationary cycle becomes increasingly long.
 A detailed analysis shows~\cite{Kree2026} that the duration time of non-propelling
states diverges $\tau*\sim(s-s^*)^{-1/2}$, as the transition is
approached from above. This critical slowing down is observed as
intermittent behavior in the displacement (see.Fig.\ref{run}): regions of
phase locked, non-winding oscillations persist for long times and
are interrupted by phase slip events that generate propulsion. The mean swimming velocity can be shown to increase continuously with a square root singularity above $s^*$.


 Within the propelling regime, the relationship between the deformation
cycle and the external forcing becomes more subtle. During one period
of driving, the shape cycle need not complete exactly one
revolution. Instead, the two cycles may synchronize in different ways,
or even may remain incommensurate.

To characterize the long-term progress along the shape cycle, we
introduce the lifted angle $\Phi(t)\in\mathbb{R}$, which records the
total advance around the circle, including the number of completed
revolutions. The original phase variable is recovered from
$
\Psi=\Phi \;{\rm mod}\; 2\pi .
$
After an initial transient, the phase advance per forcing cycle approaches a constant. This motivates the definition of the winding number
\begin{equation}\label{averageU}
  \rho=\lim_{n\to\infty}\frac{\Phi(nT)-\Phi(0)}{2\pi n}=\frac{\Delta\Phi}{2\pi}.
\end{equation}
The propulsion speed can be expressed in terms of the lifted angle as
\begin{equation}
U= {\tilde{C}}_2\dot{\Phi}+\frac{\tilde{B}_2}{2}\frac{d}{dt}\sin{2\Phi}.
\end{equation}
The synchronization properties of the shape dynamics are directly reflected in the swimming speed. In particular, the long time average
\begin{equation}\label{long_time_U}
\begin{aligned}
 <U> &=\lim_{n\to\infty} \tilde{C}_2\frac{1} {nT}\int_0^{nT}dt^{\prime} \dot{\Phi} =\\
&=\tilde{C}_2\Delta\Phi /T=\tilde{C}_2\omega\rho
\end{aligned}
\end{equation}
is completely determined by the winding number.

\begin{figure}[htbp]
    \centering
    \stackinset{l}{1pt}{t}{-8pt}{\textbf{(a)}}{
    \includegraphics[width=0.35\textwidth]{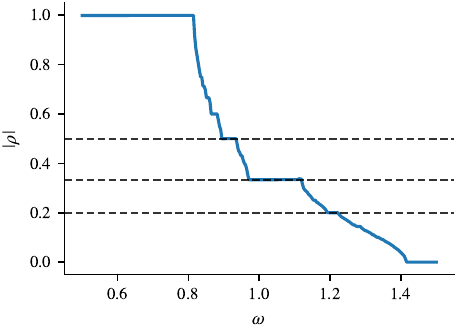}}\\
    \
    \stackinset{l}{1pt}{t}{-10pt}{\textbf{(b)}}{
    \includegraphics[width=0.35\textwidth]
   {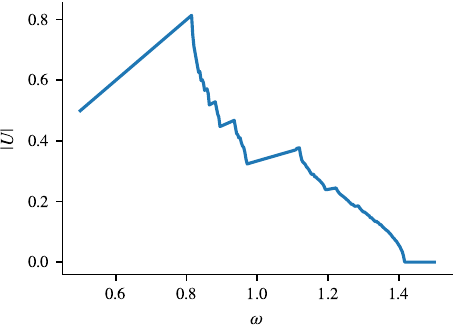}}    
    \caption{\label{Arnold2}
Synchronization and propulsion in the two-mode system. (a):
winding number $\rho$ as a function of driving frequency. Rational
plateaus correspond to synchronized states. (b): mean
propulsion velocity. Since $\langle U\rangle\propto\omega\rho$, each
plateau produces an interval of linear growth of the swimming
speed, while transitions between synchronization plateaus give rise to
local extrema. Parameter: $\hat s=10\sqrt{40}$ $\alpha=5/\sqrt{14}$, $\delta= 0.6\pi$ .}
\end{figure}

Of particular interest are rational values of $\rho=p/q$,
For example, for $\rho=1/3$ 
the vesicle needs $3$ forcing cycles for a complete
rounding of the circle, i.e. a phase change by $2\pi$. 
More generally, $\rho=p/q$ implies that the full dynamics repeats after $q$ forcing periods, during which the state advances by $p$ complete revolutions around the shape cycle. 
Such commensurate relations between the deformation cycle and the forcing cycle persist over finite parameter intervals.
These regions give rise to plateaus in the winding
$\rho$, as clearly seen at $\rho=1/3, 1/2, 2/3$ in the plot of $\rho$
versus frequency (Fig.\ref{Arnold2}).

According to Eq.~(\ref{long_time_U}), the synchronization structure is
directly reflected in the swimming speed. At low frequencies, the shape
cycle follows the driving and the mean propulsion velocity increases
approximately linearly with frequency. At high frequencies the forcing
is effectively averaged out and propulsion disappears. Between these
limits, partially synchronized states give rise to local maxima in the
mean swimming speed, reflecting the generalized phase locked states.
In the language of nonlinear dynamics, these regions are known as
Arnold tongues and will be analyzed in more detail elsewhere \cite{Kree2026}.



\subsection{Three-mode driving}

The inclusion of a third adjacent mode yields quantitatively new
dynamics: while for two-mode driving the motion in shape space is
restricted to a circle and thus one-dimensional, three-mode driving
corresponds to motion on a sphere and is thus two-dimensional,
leading to qualitatively different dynamics.

It is straightforward to extend the driving of Eq.(\ref{two_modes}) to three modes
\begin{align}\label{three_modes}
  F_2^{act}&=s\cos(\omega t),\quad F_3^{act}=\alpha_3 s\cos(\omega t+\delta_3),\nonumber\\
  F_4^{act}&=\alpha_4 s\cos(\omega t+\delta_4),
\end{align}
The constraint of constant area restricts the space of deformations: the rescaled amplitudes $q_l=f_l\sqrt{w_l/\Delta}$ lie on the unit sphere
\begin{equation}
   1=q_2^2+q_3^2+q_4^2,
\end{equation}
so that the dynamic evolution in the shape space is characterized by two angles.
The passive system exhibits fixed points $q_l^*=\pm 1$. Linear stability
analysis reveals that $q_2^*$ is stable, $q_3^*$ is a saddle point and
$q_4^*$ is unstable.

\begin{figure}
\centering
 \stackinset{l}{6pt}{t}{-12pt}{\textbf{(a)}}{\includegraphics[width=0.17\textwidth]{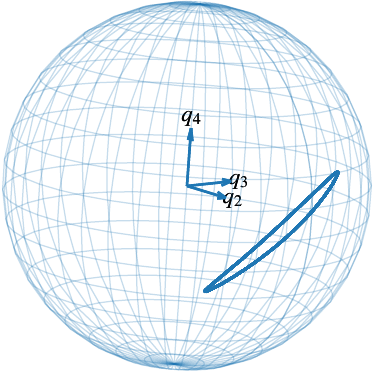}}
 \hspace{0.1\textwidth}
 \stackinset{l}{1pt}{t}{-8pt}{\textbf{(d)}}{
 \includegraphics[width=0.17\textwidth]{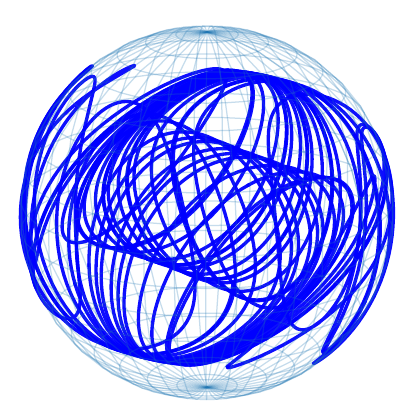}}\\ 

 \vspace{5pt}
 \centering
  \stackinset{l}{-2pt}{t}{-8pt}{\textbf{(b)}}{
 \includegraphics[width=0.15\textwidth]{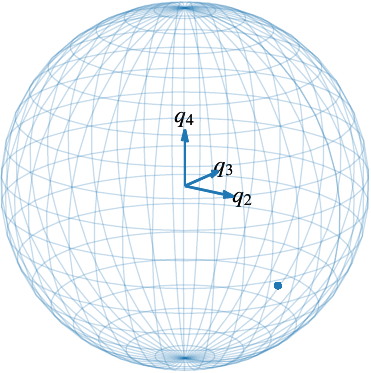}} 
 \hspace{0.1\textwidth} 
 \stackinset{l}{1pt}{t}{-8pt}{\textbf{(e)}}{
 \includegraphics[width=0.15\textwidth]{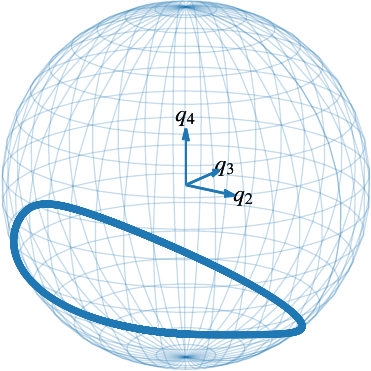}}\\ 

 \vspace{5pt}
 \stackinset{l}{15pt}{t}{-15pt}{\textbf{(c)}}{
 \includegraphics[width=0.22\textwidth]{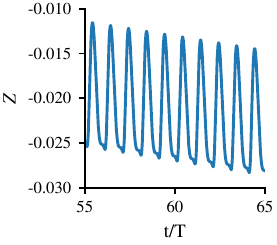}}
 \hfill
 \stackinset{l}{22pt}{t}{-15pt}{\textbf{(f)}}{
 \includegraphics[width=0.22\textwidth]{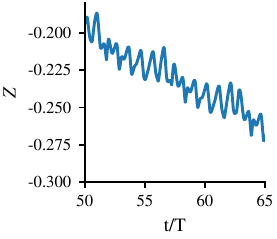}}\\
 \caption{\label{3mode_strong}
Shape dynamics and propulsion in the three-mode system. Left column, (a)--(c): periodic deformation cycle for weak forcing ($\hat{s}=5\sqrt{40}$). Right column, (d)--(f): quasiperiodic dynamics at intermediate forcing ($\hat s = 7.75\sqrt{40}$). Panels (a) and (d) show the trajectory on the shape sphere. Sampling once every forcing period yields the corresponding stroboscopic maps in (b) and (e). The resulting displacements are shown in (c) and (f). While periodic shape cycles lead to regular propulsion, quasiperiodic shape dynamics gives rise to irregular motion. Parameters: $\omega=1.48$, $\delta_3=1.55$, $\delta_4=1.0$, $\alpha_3=\sqrt{7/2}$, $\alpha_4=\sqrt{63/10}$.}

\end{figure}

An important difference from the two-mode system appears already for
weak forcing, where the dynamics remains synchronized with the
external forcing and the deformation cycle is periodic. This regime is
illustrated by the left column of Fig.~\ref{3mode_strong}. The upper
panel shows the trajectory in the shape space. Unlike the two-mode
system, where the trajectory is restricted to a circle,  the oscillations on the sphere always enclose a finite area.
The periodic nature of the dynamics is reflected in the middle panel,
which shows the stroboscopic representation. Sampling the system once
for every forcing period always returns the same point, demonstrating that
the deformation cycle repeats after one period of driving.
The resulting motion is shown in the lower panel. The
displacement increases in time, indicating steady swimming.
In contrast to the two-mode system, where weak forcing only produces
bounded oscillations without net motion, three-mode driving gives rise
to finite propulsion even for arbitrarily small activity. Movies
illustrating the deformation cycle and the resulting displacement are
provided in the supplementary material.

For increasing activity, the dynamics undergoes a transition to
quasiperiodic motion, illustrated by the right column of
Fig.~\ref{3mode_strong}. The upper panel shows that the deformation
cycle no longer closes. Instead, the trajectory explores a
two-dimensional region of shape space and gradually fills it densely.
Correspondingly, the stroboscopic representation no longer collapses to a single point but forms a closed curve, indicating that the
shape dynamics is no longer periodic with the forcing period.
The resulting displacement is shown in the lower panel. In contrast to
the regular motion observed for weak activity, the displacement now
increases irregularly over time, reflecting the quasiperiodic sequence
of shape changes. 
For very strong activity, synchronization with the external forcing is
restored. The deformation cycle again becomes periodic, and the motion
approaches a regular swimming state. Animations illustrating the different regimes are provided in
the Supplementary Videos S1-S4 in the Electronic Supplementary Material.


\section{Discussion}
\label{discussion}
The present model describes active tractions acting on a fluid membrane and is therefore largely independent of their microscopic origin. It applies equally to activity transmitted from the cortex and to active membrane properties, such as spontaneous curvature or bending rigidity. We therefore compare the characteristic time and force scales of the model with experimentally observed active processes.

Treadmilling~\cite{Vallotton2004,Bugyi2010} has long been considered an essential ingredient for
change in cell shape. Typical timescales for this process are
$\tau_{treadmilling}\sim 10 ~\mathrm{s}$. More recently, actin travelling
waves ~\cite{Weiner2007PLoSBiol,AllardMogilner2013,Bement2015PNAS} have been observed, giving rise to wavy protrusions. The
associated timescales vary in the range
$0.4 \leq\tau_{wave}\leq 4 ~\mathrm{s}$. Actin-independent driving
mechanisms are associated with membrane properties, caused by
integrated proteins. An example are MIN oscillations~\cite{Litschel2018,Christ2021} involving longer
timescales of the order of $100 ~\mathrm{s}$. The experimentally observed timescales span almost two orders of magnitude. 
One characteristic timescale of our model is set by the bending
stiffness $\kappa$ and the viscosity $\eta$ of the external medium,
according to $\tau\sim\eta a^3/\kappa $.  Using typical values
$\kappa \sim 10^{-19}$~J, $\eta \sim 10^{-3}$~Pa\,s, and
$a \sim 5$--$10~\mu$m for the vesicle radius,
one obtains
\begin{equation}
\tau \sim 1\text{--}10~\mathrm{s}.
\end{equation}
This is comparable to the timescale of actin activity, but shorter than that of protein oscillations.

The characteristic bending traction associated with a deformation mode
$\ell$ and amplitude $f_\ell$ scales as
\begin{equation}
F_b \sim \frac{\kappa}{a^3} A_\ell f_\ell,
\qquad A_\ell = O(10\text{--}10^2).
\end{equation}
Using a typical value $f_\ell \sim 10^{-2}\text{--}10^{-1},$
we obtain
\begin{equation}
F_b \sim 10^{-4}\text{--}10^{-2}\,\mathrm{Pa}
\qquad (\ell=2\text{--}4).
\end{equation}
A spatially varying spontaneous curvature $C=C_0 + C_1(\theta,t)$ produces 
traction of order
\begin{equation}
F_C \sim \frac{\kappa C_1}{a^2}.
\end{equation}
Curvature amplitudes as small as
\begin{equation}
C_1 \sim 10^{-3}\text{--}10^{-2}\,\mu\mathrm{m}^{-1}
\end{equation}
already generate tractions of the same order as the bending tractions. Similarly, 
modest rigidity contrasts $\delta\kappa/\kappa_0 \sim 0.01\text{--}0.1$ yield active tractions of comparable magnitude.
A pressure difference acts directly as normal traction. For lipid vesicles, pressure differences are typically set either by
membrane tension via the Laplace law or by
osmotic imbalances. Typical values
\begin{equation}
\Delta p \sim 10^{-1}\text{--}10^{1}~{\rm Pa}
\end{equation}
exceed the bending traction scale by several orders of
magnitude. However, the unifom pressure difference just fixes the total volume of the vesicle. Only nonuniform pressure differences can actively drive the system. Inhomogeneities of a percent are comparable to bending tractions and thus can drive deformation and propulsion.
Actomyosin activity generates stresses of order
\begin{equation}
F_{\mathrm{cortex}} \sim 1\text{--}10\,\mathrm{Pa},
\end{equation}
in lamellipodia of crawling cells~\cite{Laurent2005,Radmacher2006}. 
Compared to bending
\begin{equation}
  \frac{F_{\mathrm{cortex}}}{F_b}
  \sim 10^2\text{--}10^5,
\end{equation}
raw cortical stresses are thus much stronger and would tear the
membrane of a swimming vesicle apart. For crawling cells, adhesive
forces are presumably essential to resist such strong cortical stresses.

Direct experimental measurements of non-adhesive amoeboid swimming
provide a useful reference scale for the velocities predicted by the
present theory. In particular, trajectory analysis of freely swimming
Dictyostelium cells yields a characteristic propulsion speed of
\begin{equation}
U \approx 3~\mu{\rm m/min}
\approx 0.05~\mu{\rm m/s},
\end{equation}
as obtained from persistent random walk fits to cell trajectories
\cite{VanHaastert2011PLoS}. Values of $U$ in the range $(1\cdots 5)\, \mu {\rm m/min}$ have been observed for a variety of cells~\cite{Barry,ONEILL2018,Bodenschatz,AOUN2020}.
These values lie within the range predicted by the geometric scaling
estimate,
\begin{equation}
\bar U \sim 10^{-3}\text{--}10^{-1}~\mu{\rm m/s},
\end{equation}
derived above for quasi-spherical shape cycles. 
Of course, the comparison should be viewed with caution,  since the present model neglects many aspects of amoeboid locomotion. Nevertheless, it suggests that cyclic shape changes alone can generate swimming velocities of biologically relevant magnitude.

Taken together, the characteristic time and force scales of the model are compatible with experimentally observed values, and the predicted swimming speeds fall into the range measured for freely swimming cells.
 Experiments which simultaneously measure shape fluctuations and
propulsion velocities are rare and restricted to two dimensions. An
example is the experiments on lymphocytes~\cite{AOUN2020}, which are
, however, interpreted in terms of paddling by transmembrane proteins.
Other promising candidates for simultaneous measurements of shape and
propulsion are GUVs.  Keber et al.~\cite{Keber2014Science} study
active nematic vesicles with a few time persistent nematic defects,
causing surface flow. If the vesicles are deflated, they deform and
become motile. Another candidate is phase separated Janus
vesicles~\cite{Willems2025SoftMatter}, which are driven externally
with an AC electric field. Self-propulsion is observed for a certain
frequency range and the authors measure the propulsion velocity as a
function of driving strength and frequency. The motion resembles run-and-tumble,
but with  hardly any deformation of the GUVs.

\section{Outlook}
\label{outlook}
Our work can be extended in several directions. We have so far restricted ourselves to a few low-order modes ($\ell=2,3,4$), describing deformations on the scale of the vesicle. Including higher modes would allow for more localized protrusions and shape changes and may uncover additional deformation cycles and swimming states.

We have also limited our analysis to forcing with a single frequency. More complex frequency spectra may lead to additional synchronization phenomena. Fluctuating or stochastic components of the activity may likewise produce noise-induced transitions and intermittent switching between different motility states.

Throughout this paper, we varied only the strength and frequency of the forcing. Other parameters, such as membrane stiffness or the viscosity contrast between the inner and outer fluids, constitute additional parameters that may substantially affect the deformation cycles and swimming velocities.

We have restricted ourselves to axisymmetric shapes. Extending the formalism to general spherical harmonics $f_{\ell,m}$ is straightforward and extends the shape space from a circle or sphere to a much larger manifold. Such a description would allow turning, helical trajectories, spontaneous polarization, and other forms of motion that are inaccessible in the axisymmetric case.

The present approach establishes a link between microscopic active processes and macroscopic swimming behavior. It should therefore provide a useful framework for more detailed descriptions that incorporate internal biochemical activity and feedback between membrane mechanics, active stresses, and cortical flows. Such extensions may help clarify the relation between amoeboid swimming of micro-organisms and autonomously driven vesicles.

\newpage

\begin{appendices}

\section{Propulsion by deformation}
\label{appendixA}
All calculations will be performed in the CoD frame, because it is in this frame that the
deformations $f(\theta,t)$ are defined (see
Eq.\ref{eq:paramdeformation}). In the CoD frame, the external flow field
$\bm{v}(\bm{r},t)$ obeys the boundary condition
\begin{equation}
\label{eq:BCinfinity}
    \lim_{r\to\infty}\bm{v}^+(\bm{r},t)=-\bm{U} 
\end{equation}
at infinity. The swimming velocity $\bm{U}$ is defined here as the velocity of the center of deformation, and it points  in the z-direction, due to the uniaxial symmetry, i.e. $\bm{U}=U\bm{e}_z$. 

We expand all fields in Legendre polynomials, $P_l(\theta)$, adequate for the axisymmetric case. For the external and internal flow fields, the expansion is explicitly given by

\begin{align}
  \label{eq:vrexp}
    v_r^+(r,\theta)&=\sum_{l\geq 1}\Big(-\frac{a_l}{r^{l+2}}+
                   \frac{b_l}{r^l}\Big)(l+1)P_l(\theta)-UP_1(\theta),\\
    \label{eq:vtexp}
  v_{\theta}^+(r,\theta)&=\sum_{l\geq 1}\Big(\frac{a_l}{r^{l+2}}-\frac{(l-2)b_l}{lr^l}\Big)P_l^{\prime}(\theta)-UP_1^{\prime}(\theta),\\
\label{eq:Vrexp}
    v_r^-(r,\theta)&=\sum_{l\geq 1}\big( c_l r^{l-1}+
                   {d_l} r^{l+1}\big) lP_l(\theta)-UP_1(\theta)\\
\label{eq:Vtexp}
  v_{\theta}^-(r,\theta)&=\sum_{l\geq 1}\big(c_l r^{l-1}+\frac{(l+3)}{l+1}
                        {d_l}r^{l+1}\big)P_l^{\prime}(\theta)-UP_1^{\prime}(\theta).      
\end{align}
The boundary conditions Eqs. (\ref{eq:inextensibility},\ref{eq:BCV2}) 
determine the coefficients for $l \geq 2$ in terms of the deformation $f(\theta,t)$
\begin{align}
  a_l^{(1)}&=\frac{l}{2(l+1)}\dot{f}_l,\quad b_l^{(1)}=\frac{l+2}{2(l+1)}\dot{f}_l,\nonumber\\
   c_l^{(1)}&=\frac{l+1}{2l}\dot{f}_l,\qquad d_l^{(1)}=\frac{l-1}{2l}\dot{f}_l.
\end{align}

The $l=1$ mode requires special treatment, because
in the CODF, the shape remains
spherical and 
$
f_1=0.
$
Let us consider the calculation of the four
coefficients,
$
a_1,\, b_1,\, c_1,\, d_1,
$
and the velocity $U$ in detail.
The kinematic boundary condition Eq. (\ref{eq:BCV2}) and the inextensibility Eq. (\ref{eq:inextensibility}) imply
\begin{align}
-\frac{2a_1}{a^3}+\frac{2b_1}{a}-U&=0,
\\
\frac{a_1}{a^3}+\frac{b_1}{a}-U&=0,
\\
c_1+d_1a^2-U&=0,
\\
c_1+2d_1a^2-U&=0.
\end{align}
Hence,
\begin{equation}
a_1=\frac14Ua^3,
\qquad
b_1=\frac34Ua,
\qquad
c_1=U,
\qquad
d_1=0.
\end{equation}
The internal flow represents a rigid translation of the spherical vesicle. For an autonomous system the total momentum is conserved and hence the $1/r$ contribution in Eq. (\ref{eq:vrexp}) vanishes, which implies $U=0$.

A non-zero external force will generate a finite velocity $U$, which is obtained from the force balance. We consider arbitrary $l=1$ tractions
\begin{align}
F_r(\theta)&=F_{r,1}P_1,
\\
F_\theta(\theta)&=F_{\theta,1}P_1',
\end{align}
The corresponding total force in the $z$ direction is
\begin{equation}
   F_z 
=
\frac{4\pi a^2}{3}
\left(
F_{r,1}
+
2F_{\theta,1}
\right). 
\label{eq:total_force}
\end{equation}

The normal and tangential force balances, Eqs. (\ref{eq:force1_main}) and (\ref{eq:force2_main}), must include the surface tension $\gamma_1$ to ensure the inextensibility. 
\begin{align}
-\frac{3\eta^+U}{2a}
+\frac{2\gamma_1}{a}
+F_{r,1}
&=0,
\\
-\frac{3\eta^+U}{2a}
-\frac{\gamma_1}{a}
+F_{\theta,1}
&=0.
\end{align}
Eliminating $\gamma_1$ and using Eq. (\ref{eq:total_force}) yields
\begin{equation}
\begin{aligned}
U &=
\frac{2a}{9\eta^+}
\left(
F_{r,1}^{act}
+
2F_{\theta,1}^{act}
\right)
 = \frac{F_z}{6\pi a \eta^+} 
\end{aligned}
\end{equation}
Thus, if a net external force is present, the spherical inextensible
vesicle moves exactly like a rigid sphere obeying Stokes' law.

For an autonomous swimmer, the total force must vanish and $U=0$. 
The polar traction is then compensated entirely by the $l=1$
component of the tension field,
\begin{equation}
\gamma_1
=
aF_{\theta,1}^{act}
=
-\frac{a}{2}F_{r,1}^{act}.
\end{equation}
Consequently, autonomous propulsion cannot occur in linear order.
The net swimming arises only through the nonlinear coupling between
the deformation modes with $l \ge2$.

A finite propulsion velocity results in second order in perturbation theory in small $\epsilon$. It was calculated in Ref. \cite{Kree2025}
\begin{align}
\label{eq:Uparticle_appendix}
  U &= \sum_{l\geq 2}\big( R_l f_l\dot{f}_{l+1} +
      S_l  f_{l+1}\dot{f}_{l}\big)\quad \text{with}\\
    R_l &= -\frac{l^3 + 4l^2 +10l +3}{D_l},\\
  S_l &= -\frac{(l+2)(l^2+4)}{D_l},\\
  D_l &= (2l+1)(2l+3).
\end{align}
It is useful to separate from $U$ a part that can be written as a
time derivative, which does not contribute to the time averaged
propulsion velocity, given periodic deformations:
\begin{align}
\label{eq:Uparticle2}
  U &= \sum_{l\geq 2} C_l\big( f_l\dot{f}_{l+1} -f_{l+1}\dot{f}_{l}\big)
+\sum_{l\geq 2}B_l\frac{d}{dt} f_l\dot{f}_{l+1} 
      \quad \text{with}\\
    C_l &= -\frac{2l^2 +6l-5}{2D_l},\\
  B_l &= -\frac{2l^3+6l^2+14l+11}{2D_l}.
\end{align}

\section{Shape deformation by force balance}
\label{appendixB}

The above forces determine the traction jumps across the membrane according to
\begin{align}\label{eq:force_balance2}
 |[\sigma_{rr}]|=\sigma_{rr}^+-\sigma_{rr}^-&=-|[p]|=\mb{F}\cdot\mb{e}_r\nonumber\\
 |[\sigma_{r\theta}]|=\sigma_{r\theta}^+-\sigma_{r\theta}^-&=\mb{F}\cdot\mb{e}_\theta.
\end{align}
The flow is continuous across the membrane, so the velocity in the interior of the vesicle $\mb{v}^-$ can be calculated from $\mb{v}^-=\mb{v}^+$ on the surface $S$.

The time dependent deformation due to active driving is computed from
the force balance equations~(\ref{eq:force_balance1}).  We expand the
surface tension and the active tractions in Legendre polynomials
\begin{align}
  \label{eq:deltaC}
  F_r^{act}(\theta)&=\sum_{l\geq 1}F_{r,l}^{act}P_l(\theta),\\
                     F_{\theta}^{act}(\theta)&=\sum_{l\geq 1}F_{\theta,l}^{act}P_l^{\prime}(\theta),\\
    \label{eq:gamma}
  \gamma(\theta)&=\gamma_0+\sum_{l\geq 1}\gamma_lP_l(\theta)
                  \end{align}
The left sides of equations~(\ref{eq:force_balance2}) are computed with the flow fields from the previous section to yield
\begin{align}\label{eq:hydrodynamics}
  |[\sigma_{rr}]|&=-\sum_{l\geq 2}N_l\dot{f}_{l}P_l(\theta)\nonumber\\
N_l&= \eta\frac{(l+2)(2l-1)}{a(l+1)} +\lambda\frac{(l-1)(2l+3)}{al}\\
  |[\sigma_{r\theta}]|&=-\sum_{l\geq 2}T_l\dot{f}_{l}P_l^{\prime}(\theta)\nonumber\\
T_l&= \frac{\eta(l+2)+\lambda(l-1)}{al(l+1)} 
\end{align}

We substitute the expansions (\ref{eq:deltaC},\ref{eq:gamma}) into Eq.(\ref{eq:force_balance2})and decompose the force into normal and tangential components to obtain
 two equations of motion for $\dot{f}_l$:
\begin{align}
  \label{eq:force1}
  -N_l\dot{f}_l&=g_l\Big(\frac{\kappa}{a^3}g_l+\frac{\gamma_0}{a}\Big)f_l
 +\frac{2}{a}\gamma_l +F_{r,l}^{act}\\
    \label{eq:force2}
  -T_l\dot{f}_l &=-\frac{\gamma_l}{a}+F_{\theta,l}^{act}
\end{align}
with $g_l=l(l+1)-2$.


Eq.(\ref{eq:force2}) determines $\gamma_l$, which is substituted in Eq.(\ref{eq:force1}) to obtain the evolution equation for the
deformation, driven by active tractions:
\begin{equation}
  -(N_l+2T_l)\dot{f}_{l}=g_l(\frac{\gamma_0}{a}+\frac{\kappa}{a^3}g_l)f_l
  +F_{r,l}^{act}+2F_{\theta,l}^{act}.
\end{equation}

The constraint of constant area (\ref{eq:area_constraint}) implies
\begin{equation}
  \dot{\Delta}=0=\sum_l\frac{g_l}{2l+1}f_l\dot{f}_l.
\end{equation}
and is enforced by choosing $\gamma_0$ appropriately:
\begin{align}
  \frac{\gamma_0}{a}&\sum_{l \leq 2}\frac{g_l^2f_l^2}{(2l+1)(N_l+2T_l)}\nonumber\\
  &=-
  \sum_{l \leq 2}\frac{f_lg_l}{(2l+1)(N_l+2T_l)}\Big(\frac{\kappa}{a^3}g_l^2f_l
  +F_{r,l}^{act}+2F_{\theta,l}^{act}  \Big)
\end{align}

\section{Active and passive bending forces}
\label{appendixC}

We need to derive the bending forces from the Helfrich Hamiltonian. In
principle, this is well known and given in the literature~\cite{Deserno2015}; nevertheless, we
 present a simple case here to make the presentation more
readable. We start from the Helfrich energy
\begin{equation}
 {\cal{F}}_H=\int_S dA \frac{\kappa}{2} (H-C)^2 
\end{equation}
and consider an inhomogeneous spontaneous curvature,
$C=2+\epsilon\delta C$, as the only driving mechanism. Other possible
driving mechanisms, such as an inhomogeneous bending modulus $\kappa$
or a different spontaneous homogeneous curvature ($C_0\neq 2$) are not
considered in this Appendix, but are included in the main text. To
obtain the force from the Helfrich energy, we need the first variation
of the free energy with respect to radial and tangential variations. We use the expansion of the mean curvature, given
in Eq.(\ref{mean_curvature_expansion}), to expand the free energy density
\begin{equation}
(H-C)^2 =\frac{\epsilon^2}{a^2}(2+\Delta_{\Omega})f\Big((2+\Delta_{\Omega})f+2\delta C\Big) +{\cal{O}}(\epsilon^3),\nonumber
\end{equation}
where $\Delta_{\Omega}$ denotes the Laplacian on the unit sphere. We first consider radial variations $\mb{x}=a(1+f)\mb{e}_r+\delta f\mb{e}_r$ and linearize the free energy in the variation
\begin{align}
  \frac{2a^2}{\kappa} \delta {\cal{F}}_H&=\int_S dA\Big(2(2+\Delta_{\Omega})f+2\delta C\Big)                                       
                                       \nonumber\\
  &
    (2+\Delta_{\Omega})\delta f . 
\end{align}
Given the expansion in $\epsilon$, the area element is simply given by $dA=\sin{\theta} \,d\theta \,d\phi$.
In the next step, we use an integration by parts twice, in order to remove the Laplacian from $\delta f$. The boundary terms vanish, so that we arrive at
\begin{equation}
  \frac{\delta {\cal{F}}_H}{\delta f}=\frac{\kappa}{a^2} (2+\Delta_{\Omega})\Big(
(2+\Delta_{\Omega})f+\delta C  \Big).
\end{equation}
The variation in the tangential direction vanishes, which is known as
invariant parameterization and can also be shown explicitly along the
lines above. We have thus derived the normal force due to an inhomogeneous curvature, as displayed in Eq.(\ref{eq:normal_force}) with
\begin{equation}
F_r^{act}=\frac{\kappa}{a^2}(2+\Delta_{\Omega})\delta C.
\end{equation}
  
\section{Differential equation for $\Psi(t)$}
\label{appendixD}

We want to derive an equation of motion for the angle $\Psi$, starting
from Eq.(\ref{evolution_f}), and rewrite it in terms of $q_l=f_l\sqrt{w_l/\Delta}$
\begin{equation}
  \dot{q}_l=-x_l(\frac{\gamma_0}{a}+\beta_l)q_l-Z_l.
\end{equation}
Here we have introduced the abbreviations
\begin{equation}
  x_l=\frac{g_l}{N_l+2T_l},\quad \beta_l=\frac{\kappa g_l}{a^3},
  \quad Z_l=\sqrt{\frac{w_l}{\Delta}}\;\frac{F_l^{act}}{N_l+2T_l}.
\end{equation}

Specialising to two-mode driving, we use polar coordinates
$q_2=\cos{\Psi}$ and $q_3= \sin{\Psi}$, implying
\begin{equation}
  \dot{\Psi}=\dot{q}_3q_2-\dot{q}_2q_3.
\end{equation}
Substituting the equation of motion for $ \dot{q}_l$, we arrive at
 \begin{align}\label{eq:phase}
    \dot{\Psi}=q_2q_3\Big(x_2(\frac{\gamma_0}{a}+\beta_2)-x_3(\frac{\gamma_0}{a}+\beta_3)
   \Big)-q_2Z_3+q_3Z_2.
 \end{align}
 The last step in the calculation is the elimination of $\gamma_0$, which ensures the constant excess area: $\dot{q}_2q_2+\dot{q}_3q_3=0$. Solving this equation for $\gamma_0$ yields
 \begin{equation}\label{gamma0}
 \mu(\Psi)\frac{\gamma_0}{a}=-q_2(Z_2+x_2\beta_2q_2)-q_3(Z_3+x_3\beta_3q_3),
\end{equation}
with $\mu(\Psi)=x_2\cos^2{\Psi}+x_3\sin^2{\Psi}$.
We substitute the result for $\gamma_0$ in Eq.(\ref{eq:phase}) and perform some algebraic manipulations to finally obtain the equation of motion for the angle $\Psi$:
\begin{align}
  \mu(\Psi) \dot{\Psi}=-&\frac{x_2x_3}{2}(\beta_3-\beta_2)\sin{(2\Psi)}\nonumber\\
  +&x_3Z_2\sin{\psi}
-x_2Z_3\cos{\psi}
\end{align}
Recalling the scaling, $Z_l\propto F_l/\sqrt{\Delta}$, we see that the effective forcing strength is $s/\Delta$. The passive system ($s=0$) has 2 stable fixed points $\Psi=0,\pi$, as mentioned in the main text. Without forcing, the dynamics is purely relaxational with rate  $\Omega_0=x_2x_3(\beta_3-\beta_2)/2>0$.
\end{appendices}

\section*{Data availability}

The data supporting the findings of this study are available within the article and its Electronic Supplementary Material. Additional numerical data and computer code are available from the corresponding author upon reasonable request.

\bibliography{active_vesicle}

@article{Deserno2015,
  author  = {Deserno, M.},
  title   = {Fluid lipid membranes: from differential geometry to
                  curvature stresse},
  journal = {Chemistry and Physics of Lipids},
  year    = {2015},
  volume  = {185},
  number  = {},
  pages   = {031918},
 }

@article{Lim2013,
author = {Fong Yin Lim and Yen Ling Koon and Keng-Hwee Chiam},
title = {A computational model of amoeboid cell migration},
journal = {Computer Methods in Biomechanics and Biomedical Engineering},
volume = {16},
number = {10},
pages = {1085--1095},
year = {2013},
publisher = {Taylor \& Francis},
doi = {10.1080/10255842.2012.757598}
}

@article{Holderer2013,
author = {Holderer, O. and Frielinghaus, H. and Monkenbusch, M. and Klostermann, M. and Sottmann, T. and Richter, D.},
title = {Experimental determination of bending modulus and saddle splay modulus in bicontinuous microemulsions},
journal = {Soft matter},
volume = {9},
number = {},
pages = {2308},
year = {2013},
}

@article{Leermakers2013,
author = {Holderer, F.},
title = {Direct evaluation of the saddle splay modulus of a
                  liquid-liquid interface using the classical mean
                  field lattice model},
journal = {J. Chem. Physics},
volume = 138,
pages = 124103,
year = 2013,
}

@article{McMahon2005,
author = {McMahon, H. T. and Gallop, J. L.},
title = {Membrane curvature and mechanisms of dynamics cell membrane
                  remodelling},
journal = {Nature},
volume = 438,
pages = 590,
year = 2005,
}

@article{Dimova2014,
author = {Dimova, R.},
title = {Recent developments in the field of bending rigidity
                  measuerments of membranes},
journal = {Adv. Colloid and Interface Science},
volume = 208,
pages = 225,
year = 2014,
}

@article{HuBriguglioDeserno2012,
  title   = {Determining the Gaussian Curvature Modulus of Lipid
                  Membranes in Simulations},
  author  = {Hu, Mingjie and Briguglio, Joseph J. and Deserno, Markus},
  journal = {Biophysical Journal},
  volume  = {102},
  number  = {6},
  pages   = {1403--1410},
  year    = {2012},
  doi     = {10.1016/j.bpj.2012.02.013}
}

@article{Bergert,
  author = {M. Bergert and A. Erzberger and R. A. Desai and I. M. Aspalter and A. C. Oates and G. Charras and G. Salbreux and E. K. Paluch},
  title = {Force transmission during adhesion-independent migration},
  year = {2015},
  journal = {Nat. cell Biol.},
  volume = {17},
  pages = {524},
}

@article{CallenJones,
  author = {A. C. Callen-Jones and V. Ruprecht and S. Wieser and C.P. Heisenberg and R. Voituriez},
  title = {Critical flow-driven shapes of nonadherent cells},
  year = {2016},
  journal = {Phys. Rev. Lett.},
  volume = {116},
  pages = {028102},
}

@article{Paluch2016,
  author = {E. K. Paluch and I. M. Aspalter and M. Sixt},
  title = {Focal adhesion-independent cell migration},
  year = {2016},
  journal = {Annu. Rev. Cell Dev. Biol.},
  volume = {32},
  pages = {469},
}

@article{Andrieu,
  author = {Cyril Andrieu and Bren Hunyi Lee and Anna Franz},
  title = {Cell deformations generated by stochastic actomyosin waves in vivo random-walk swimming migration}}

@article{Othmer2019,
  author = {H. G. Othmer},
  title = {Eukaryotic cell dynamics from crawling to swimming},
  year = {2019},
  journal = {Wiley Interdiscip Rev Comput Mol Sci.},
  volume = {9},
  pages = {1},
}

@article{Barry,
  author = {N. P. Barry and M. S. Bretscher},
  title = {Dictyostelium amoeba and neutophils can swim},
  year = {2010},
  journal = {PNAS},
  volume = {107},
  pages = {11376},
}

@article{ONEILL2018,
title = {Membrane Flow Drives an Adhesion-Independent Amoeboid Cell Migration Mode},
journal = {Developmental Cell},
volume = {46},
number = {1},
pages = {9-22.e4},
year = {2018},
issn = {1534-5807},
doi = {https://doi.org/10.1016/j.devcel.2018.05.029},
author = {Patrick R. O'Neill and Jean A. Castillo-Badillo and Xenia Meshik and Vani Kalyanaraman and Krystal Melgarejo and N. Gautam},
}

@article{AOUN2020,
title = {Amoeboid Swimming Is Propelled by Molecular Paddling in Lymphocytes},
journal = {Biophysical Journal},
volume = {119},
number = {6},
pages = {1157-1177},
year = {2020},
issn = {0006-3495},
doi = {https://doi.org/10.1016/j.bpj.2020.07.033},
author = {Laurene Aoun and Alexander Farutin and Nicolas Garcia-Seyda and Paulin Nègre and Mohd Suhail Rizvi and Sham Tlili and Solene Song and Xuan Luo and Martine Biarnes-Pelicot and Rémi Galland and Jean-Baptiste Sibarita and Alphée Michelot and Claire Hivroz and Salima Rafai and Marie-Pierre Valignat and Chaouqi Misbah and Olivier Theodoly},
}

@article{Bodenschatz,
  author = {A. J. Bae and E. Bodenschatz},
  title = {On the swimming of Dictyostelium amoeba},
  year = {2010},
  journal = {PNAS},
  volume = {107},
  pages = {E165},
}

@article{Jones,
  author = {A. Callen-Jones},
  title = {Self-organization in amoeboid motility},
  journal = {Frontiers in cell and and developmental biology},
  year = 2022,
  month = {October},
  volume = 10,
  pages = {},
  DOI={10.3389/fcell.2022.1000071}
}

@article{Noselli,
  author  = {Giovanni Noselli and Alfred Beran and Marino Arroyo and Antonio DeSimone},
  title   = {Swimming Euglena respond to confinement with a behavioural change enabling effective crawling},
  journal = {Nature physics},
  year    = {2019},
  volume  = {15},
  number  = {},
  pages   = {496--502},
}

@article{Garcia_Seyda,
author = {Garcia-Seyda, Nicolas and Seveau, Valentine and Manca, Fabio and Biarnes-Pelicot, Martine and Valignat, Marie-Pierre and Bajénoff, Marc and Theodoly, Olivier},
title = {Human neutrophils swim and phagocytise bacteria},
journal = {Biology of the Cell},
volume = {113},
number = {1},
pages = {28-38},
doi = {https://doi.org/10.1111/boc.202000084}
}

@article{Litschel2018,
  title={Beating vesicles: Encapsulated protein oscillations cause dynamic membrane deformations},
  author={Litschel, Thomas and Ramm, Beatrice and Maas, Roel and Heymann, Michael and Schwille, Petra},
  journal={Angewandte Chemie International Edition},
  volume={57},
  number={6},
  pages={1621--1626},
  year={2018},
  doi={10.1002/anie.201710374}
}

@article{Jones2020,
  author  = {Jones, T. (IV) and Liu, A. and Cui, B.},
  title   = {Light-Inducible Generation of Membrane Curvature in Live Cells with Engineered BAR Domain Proteins},
  journal = {ACS Synthetic Biology},
  year    = {2020},
  volume  = {9},
  number  = {4},
  pages   = {893--901},
  doi     = {10.1021/acssynbio.9b00516}
}

@article{Christ2021,
  author       = {Christ, Simon and Litschel, Thomas and Schwille, Petra and Lipowsky, Reinhard},
  title        = {Active shape oscillations of giant vesicles with cyclic closure and opening of membrane necks},
  journal      = {Soft Matter},
  year         = {2021},
  volume       = {17},
  pages        = {319-330},
  doi          = {10.1039/D0SM00790K}
}

@article{AbaurreaVelasco2019,
  author  = {Abaurrea-Velasco, Clara and Auth, Thorsten and Gompper, Gerhard},
  title   = {Vesicles with Internal Active Filaments: Self-Organized Propulsion Controls Shape, Motility, and Dynamical Response},
  journal = {New Journal of Physics},
  year    = {2019},
  volume  = {21},
  pages   = {123024},
  doi     = {10.1088/1367-2630/ab5c70}
}

@article{Paoluzzi2016,
  author  = {Paoluzzi, Matteo and Di Leonardo, Roberto and Marchetti,
                  M. Christina and Angelani, Luca},
  title   = {Shape and displacement fluctuations in soft vesicles
                  filled by active particles},
  journal = {Scientific reports},
  year    = {2016},
  volume  = {6:34146},
  pages   = {},
  doi     = {10.1038/srep34146}
}

@article{Tian2017,
  author  = {Tian, Wen-de and Gu, Yan and Guo, Yong-Kun and Chen, Kang},
  title   = {Anomolous boundary deformation induced by enclosed active particles},
  journal = {Chinese Physics B},
  year    = {2017},
  volume  = {26},
  pages   = {100502}
  }

@article{Iyer2022,
  author  = {Iyer, Priyanka and Gompper, Gerhard and Fedosov, Dmitry A.},
  title   = {Non-Equilibrium Shapes and Dynamics of Active Vesicles},
  journal = {Soft Matter},
  year    = {2022},
  volume  = {18},
  number  = {36},
  pages   = {6868--6881},
  doi     = {10.1039/D2SM00622G}
}

@article{Farutin2013,
  author  = {Farutin, Alexander and Rafai, Salima and Dysthe, Dag
                  Kristian and Duperray, Alain and Peyla, Philippe and
                  Misbah, Chaouqi},
  title   = {Amoeboid swimming: A generic self-propulsion of cells in fluids by means of membrane deformations},
  journal = {Phys. Rev. Lett.},
  year    = 2013,
  volume  = 111,
  pages   = 228102}

@article{Kree2025,
  author  = {Kree, Reiner and Zippelius, Annette},
  title   = {Amoeboid Propulsion of Active Solid Bodies, Vesicles and Droplets: A Comparison},
  journal = {Soft Matter},
  year    = {2025},
  volume  = {21},
  pages   = {4241--4255},
  doi     = {10.1039/D4SM01504E}
}

@article{Kree2026,
  author  = {Kree, Reiner and Zippelius, Annette},
  title   = {Shape dynamics and self-propulsion of area-preserving vesicles},
  journal = {preprint},
  year    = {2026},
  volume  = {},
  pages   = {},
 }

@article{Lighthill,
  author = {M. J. Lighthill},
  title = {On the squirming motion of nearly spherical deformable bodies through liquids at very small Reynolds numbers},
  year = {1952},
  journal = {Comm. Pure Appl. Math.},
  volume = {5},
  pages = {109},
}

@article{Barrio2020,
  doi = {10.1371/journal.pone.0227562},
  url = {https://doi.org},
  year = {2020},
  month = jan,
  publisher = {Public Library of Science ({PLoS})},
  volume = {15},
  number = {1},
  pages = {e0227562},
  author = {R. A. Barrio and T. Alarcon and A. Hernandez-Machado},
  title = {The dynamics of shapes of vesicle membranes with time dependent spontaneous curvature},
  journal = {{PLOS} {ONE}}
}

@article{Pernpeintner2017_Langmuir,
  author  = {Pernpeintner, C. and Frank, J. A. and Lohm{\"u}ller, T. and others},
  title   = {Light-Controlled Membrane Mechanics and Shape Transitions of Photoswitchable Lipid Vesicles},
  journal = {Langmuir},
  year    = {2017},
  volume  = {33},
  pages   = {4083--4089},
  doi     = {10.1021/acs.langmuir.7b01020}
}

@article{Pritzl2025_Review,
  author  = {Pritzl, S. D. and Morstein, J. and Pritzl, N. A. and
                  Lipfert, J. and Lohm{\"u}ller, T. and Trauner, D. H.},
  title   = {Photoswitchable phospholipids for the optical control of
                  membrane processes, protein function, and drug
                  delivery},
  journal = {Communications Materials},
  year    = {2025},
  volume  = {6:59},
  pages   = {},
  doi     = {10.1038/s43246-025-00773-8}
}

@article{Schwille2012,
  author = {Vogel, Stefan K. and Schwille, Petra},
  title = {Minimal systems to study membrane-cytoskeleton interactions},
  journal = {Current oppinion in biotechnology},
  year = {2012},
  volume = {23},
  pages = {758-765}
}

@article{Keber2014Science,
  author = {Keber, Felix C. and Loiseau, Etienne and Sanchez, Tim and DeCamp, Stephen J. and Giomi, Luca and Bowick, Mark J. and Marchetti, M. Cristina and Dogic, Zvonimir and Bausch, Andreas R.},
  title = {Topology and dynamics of active nematic vesicles},
  journal = {Science},
  year = {2014},
  volume = {345},
  pages = {1135--1139}
}

@article{Willems2025SoftMatter,
  author = {Willems, Vivien and Baron, Alexandre and Fernandez-Matos,
                  Daniel and Wolfisberg, Gianna and Baret, Jean-Christoph and
                  Dufresne, Eric and Alvarez, Laura},
  title = {Run-and-tumble dynamics of active giant vesicles},
  journal = {Soft Matter},
  year = {2025},
  volume = {21},
  pages = {6175--6185}
}

@article{Bement2015PNAS,
  author  = {Bement, William M. and Leda, Marcin and Moe, Alison M. and Kita, Alexander M. and Larson, Megan E. and Golding, Adriana E. and Pfeuti, Corinne and Su, Kun-Chun and Miller, Andrew L. and Goryachev, Andrew B. and von Dassow, George},
  title   = {Activator–inhibitor coupling between Rho signalling and actin assembly makes the cell cortex an excitable medium},
  journal = {Proceedings of the National Academy of Sciences},
  year    = {2015},
  volume  = {112},
  number  = {44},
  pages   = {E6603--E6611},
  doi     = {10.1073/pnas.1508873112}
}

@article{Vallotton2004,
author = {Pascal Vallotton  and Stephanie L. Gupton  and Clare M. Waterman-Storer  and Gaudenz Danuser },
title = {Simultaneous mapping of filamentous actin flow and turnover in migrating cells by quantitative fluorescent speckle microscopy},
journal = {Proceedings of the National Academy of Sciences},
volume = {101},
number = {26},
pages = {9660-9665},
year = {2004},
doi = {10.1073/pnas.0300552101}
}

@article{Weiner2007PLoSBiol,
  author  = {Weiner, Orion D. and Marganski, William A. and Wu, Liufang F. and Altschuler, Steven J. and Kirschner, Marc W.},
  title   = {An actin-based wave generator organizes cell motility},
  journal = {PLoS Biology},
  year    = {2007},
  volume  = {5},
  number  = {9},
  pages   = {e221},
  doi     = {10.1371/journal.pbio.0050221}
}

@article{AllardMogilner2013,
  author  = {Allard, Jun and Mogilner, Alex},
  title   = {Traveling waves in actin dynamics and cell motility},
  journal = {Current Opinion in Cell Biology},
  year    = {2013},
  volume  = {25},
  number  = {1},
  pages   = {107--115},
  doi     = {10.1016/j.ceb.2012.08.012}
}

@article{Bugyi2010,
  author  = {Bugyi, B. and Carlier, M.-F.},
  title   = {Control of actin filament treadmilling in cell motility},
  journal = {Annu Rev Biophys.},
  year    = {2010},
  volume  = {39},
  number  = {},
  pages   = {449--470},
  doi     = {10.1146/annurev-biophys-051309-103849}
}

@article{VanHaastert2011PLoS,
  author  = {Van Haastert, Peter J. M.},
  title   = {Amoeboid cells use protrusions for walking, gliding and swimming},
  journal = {PLoS ONE},
  year    = {2011},
  volume  = {6},
  number  = {11},
  pages   = {e27532},
  doi     = {10.1371/journal.pone.0027532}
}

@article{Laurent2005,
  author  = {Laurent, V. M. and Kasas, S. and Yersin, A. and Sch\"affer,
                  T. E. and Catsicas, S. and Dietler, G. and
                  Verkhovsky, A. B. and Meister, J. J.},
  title   = {Gradient of rigidity in the lamellipodia of migrating
                  cells revealed by atomic force microscopy},
  journal = {Biophys J.},
  year    = {2005},
  volume  = {89},
  number  = {1},
  pages   = {667--675},
  doi     = {10.1529/biophysj.104.052316}
}

@article{Radmacher2006,
  author  = {Prass, M. and Jacobson, K. and Mogilner, A: and Radmacher, M.},
  title   = {Direct measurement of the lamellipodial protrusive force
                  in a migrating cell},
  journal = {J Cell Biol.},
  year    = {2006},
  volume  = {174},
  number  = {},
  pages   = {767--772},
  doi     = {doi: 10.1083/jcb.200601159}
}

@article{Campbell2017,
  title = {A computational model of amoeboid cell swimming},
  author = {Campbell, Eric J. and Bagchi, Prosenjit},
  journal = {Physics of Fluids},
  volume = {29},
  number = {10},
  pages = {101902},
  year = {2017},
  doi = {10.1063/1.4990543},
  url = {https://pubs.aip.org/aip/pof/article/29/10/101902/602148/A-computational-model-of-amoeboid-cell-swimming}
}
\end{document}